\documentclass[12pt]{iopart}
\usepackage{iopams}
\usepackage{bm}
\usepackage[dvips]{graphicx}
\usepackage{color}

\newcommand{\ui}{{\rm i}}
\newcommand{\kB}{k_{\rm B}}
\newcommand{\bmr}{{\bm r}}
\newcommand{\bmq}{{\bm q}}
\newcommand{\bmk}{{\bm k}}

\newcommand{\bmK}{{\bm K}}

\newcommand{\bmS}{{\bm S}}
\newcommand{\bmH}{{\bm H}}

\newcommand{\bra}{{\langle}}
\newcommand{\ket}{\rangle} 
\newcommand{\bmsig}{{\bm \sigma}}
\newcommand{\eps}{\varepsilon}

\begin{document}

\title{Spin Seebeck effect in a simple ferromagnet near $T_{\rm c}$: A Ginzburg-Landau approach }

\author{Hiroto Adachi, Yutaka Yamamoto, and Masanori Ichioka}

\address{Research Institute for Interdisciplinary Science, Okayama University, Okayama 700-8530, Japan} 
\ead{hiroto.adachi@okayama-u.ac.jp}
\vspace{10pt}
%\begin{indented}
%\item[]February 2014
%\end{indented}

\begin{abstract}
  A time-dependent Ginzburg-Landau theory is used to examine the longitudinal spin Seebeck effect in a simple ferromagnet in the vicinity of the Curie temperature $T_{\rm c}$. It is shown analytically that the spin Seebeck effect is proportional to the magnetization near $T_{\rm c}$, whose result is in line with the previous numerical finding. It is argued that the present result can be tested experimentally using a simple magnetic system such as EuO/Pt or EuS/Pt. 
\end{abstract}

%\maketitle

\section{Introduction}
The magnonic thermal spin injection phenomenon from a magnet into the adjacent heavy metal is referred to as spin Seebeck effect (SSE)~\cite{Bauer-review12}. The SSE~\cite{Uchida08,Jaworski10,Uchida10} not only offers a concise way of creating spin currents, but also provides a good opportunity to examine the basic physics of magnonic spin transport. Investigations of the physics behind the SSE now extend to multilayer SSE~\cite{Ramos15}, time-resolved SSE~\cite{Schreier16,Kimling17}, paramagnetic and antiferromagnetic SSE~\cite{Wu15,Seki15,Wu16}, ferrimagnetic SSE near the compensation point~\cite{Gepraegs16}, and the SSE in bulk nanocomposites~\cite{Boona16}. So far, it has been understood that thermally-excited magnons play a central role in the SSE at room temperature~\cite{Xiao10,Adachi11,Adachi13}. Upon cooling, on the other hand, the contribution of long-lived phonons dragging magnons gets more and more important~\cite{Adachi10,Jaworski11,Iguchi17}. 

Recently, this magnonic senario has been challenged by an experimental finding that the SSE in yttrium iron garnet (YIG) shows a power law behavior [$\sim (T_{\rm c}-T)^3$] near $T_{\rm c}$~\cite{Uchida14}. Subsequently, an atomistic numerical simulation of Heisenberg Hamiltonian~\cite{Barker16} concluded that a different behavior [$\sim (T_{\rm c}- T)^{1/2}$] is expected if we rely on a simple magnonic picture. Therefore it is of vital importance to examine the origin of the disagreement, and a more simple analytical approach that can shed light on the underlying physics is desired. 

In this paper we focus on a simple ferromagnet composed solely of a single sublattice, and study the longitudinal SSE~\cite{Uchida10b} in the vicinity of $T_{\rm c}$. In contract to the approach of Ref.~\cite{Barker16} employing an atomistic numerical simulation, in the present work we use a time-dependent Ginzburg-Landau (TDGL) model~\cite{Ma75}. The TDGL model is considered to be a minimal model for a ferromagnet near $T_{\rm c}$, relevant for describing its relaxational dynamics~\cite{Hohenberg77}. Starting from this model we show that, if we only consider the local spin transfer process across the ferromagnet/heavy-metal interface, we recover the result of Ref.~\cite{Barker16}, i.e., the SSE signal scales with magnetization [$\sim (T_{\rm c}- T)^{1/2}$]. Furthermore, we discuss the effects of spin diffusion on the SSE and argue that the above conclusion, i.e., the SSE scales with magnetization, is unchanged. 

This paper is organized as follows. In the next section, the TDGL model is introduced and its behavior is described. In Sec.~\ref{Sec:III}, we focus on the local spin injection process and calculate the resultant SSE. In Sec.~\ref{Sec:IV}, we discuss the effects of spin diffusion through the ferromagnet. Finally in Sec.~\ref{Sec:V}, we discuss and summarize the present result. 

%%%%%%%%%%%%%%%%%%%%%%%%%%%%%%%%%
\section{Model} \label{Sec:II}
%%%%%%%%%%%%%%%%%%%%%%%%%%%%%%%%%

%%%%%%%%%%%%%%%%%%%%%%%%%%%%%%%%%%%%% 
\begin{figure}[t] 
  \begin{center}
    \includegraphics[width=5cm]{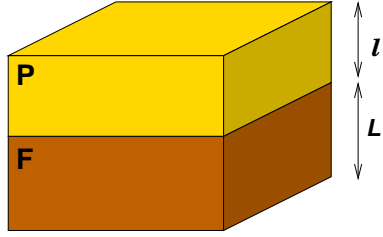}
  \end{center}
\caption{ 
Schematic view of the system studied in this paper. A paramagnetic metal $P$ with thickness $l$ is attached on top of a ferromagnet $F$ with thickness $L$. 
}
\label{fig1_adachi}
\end{figure}
%%%%%%%%%%%%%%%%%%%%%%%%%%%%%%%%%%%%% 

We consider a system as depicted in figure~\ref{fig1_adachi}, where the longitudinal SSE injects spins from a ferromagnet $F$ into the adjacent paramagnetic metal $P$. The ferromagnet is assumed to locate in the vicinity of the Curie temperature $T_{\rm c}$. The spin information in $F$ and $P$ is communicated through the $s$-$d$ interaction at the $F/P$ interface. 

We begin with the TDGL equation valid near the Curie temperature of a ferromagnet~\cite{Ma75,Ohnuma14}: 
%%%
\begin{equation}
  \frac{\partial}{\partial t} \bmS(\bmr) 
  =
  \left[ \gamma \bmH_{\rm eff}(\bmr) + \frac{J_{\rm sd}(\bmr)}{\hbar} \bmsig(\bmr) \right] \times \bmS(\bmr)
  + \big( \Gamma_0 -D_0 {\bm \nabla}^2 )\frac{\bmH_{\rm eff}(\bmr)}{\mathfrak{h}_0}
  + {\bm \xi}(\bmr), \label{eq:TDGL01}
\end{equation}
%%%
where $\bmS$ is a coarse-grained spin within an effective cell volume $v_0$ in $F$, $\gamma$ is the gyromagnetic ratio, $\Gamma_0$ is the dissipation coefficient, $D_0$ is the spin diffusion coefficient, and $\mathfrak{h}_0= \gamma \hbar/v_0$ is the unit of magnetic field. In the above equation, 
%%%
\begin{equation}
  \bmH_{\rm eff} (\bmr) = \bmH_0 - \mathfrak{h}_0^{-1} \frac{\delta F_{\rm GL}}{\delta \bmS(\bmr)}
\end{equation}
%%%
is the effective field, $\bmH_0$ is the external magnetic field, and 
%%%
\begin{equation}
  F_{\rm GL} = \eps_0 \int d^3r \; \left\{ \frac{a_{\rm GL}}{2} \bmS(\bmr)^2
  + \frac{b_{\rm GL}}{4} \bmS(\bmr)^4 
  + \frac{c_{\rm GL}}{2} \big( {\bm \nabla} \bmS(\bmr) \big)^2 \right\} 
\end{equation}
%%%
is the Ginzburg-Landau free energy of the ferromagnet, where $\eps_0= \mathfrak{h}_0^2$ is the magnetic energy density, $a_{\rm GL}= (T- T_{\rm c})/T_{\rm c}$ measures the distance from the Curie temperature, $b_{\rm GL}$ is the quartic term coefficient, and $c_{\rm GL}$ has the meaning of the square of the correlation length. Besides, effects of the $s$-$d$ interaction at the interface is described by
%%%
\begin{equation}
  J_{\rm sd}(\bmr) = J_{\rm sd} \widetilde{\rho}(\bmr), 
\end{equation}
%%%
where $\widetilde{\rho}(\bmr)= \sum_{\bmr_0} v_0 \delta(\bmr- \bmr_0)$ is the (normalized) density function of spin $\bmS$ at the interface. Finally, the last term of equation~(\ref{eq:TDGL01}) represents the effect of thermal noise, which is assumed to obey the following Gaussian ensemble~\cite{Ma75}: 
%%%
\begin{eqnarray}
  \bra \xi^i(\bmr,t) \ket &=& 0, \\
  \bra \xi^i(\bmr,t) \xi^j(\bmr',t') \ket &=& \frac{2 \kB T_F }{\eps_0} (\Gamma_0- D_0\nabla^2) 
  \delta_{i,j} \delta(\bmr-\bmr') \delta(t-t'),
  \label{eq:FDxi01}
\end{eqnarray}
%%%
where $T_F$ is the temperature of $F$.

In the paramagnetic metal $P$, the dynamics of the itinerant spin $\bmsig$ is described by the Bloch equation: 
%%%
\begin{equation}
  \frac{\partial}{\partial t} \bmsig(\bmr)  =
  \frac{J_{\rm sd}(\bmr)}{\hbar} \bmS(\bmr) \times \bmsig(\bmr)
  - \frac{1}{\tau_P} \Big( \bmsig(\bmr) - \chi_P J_{\rm sd}(\bmr) \bmS(\bmr) \Big)+ {\bm \zeta}(\bmr),
  \label{eq:Bloch01}  
\end{equation}
%%%
where $\chi_P$ is the paramagnetic susceptibility of $P$ having the dimension of energy$^{-1}$, $\tau_P$ is the relaxation time of $\bmsig$, and ${\bm \zeta}$ is the Gaussian thermal noise:
%%%
\begin{eqnarray}
  \bra \zeta^i(\bmr,t) \ket &=& 0, \\
  \bra \zeta^i(\bmr,t) \zeta^j(\bmr',t') \ket &=& \frac{2 \kB T_P \chi_P v_0}{\tau_P} 
  \delta_{i,j} \delta(\bmr-\bmr') \delta(t-t').
  \label{eq:FDzeta01}
\end{eqnarray}
%%%

Since our approach to the SSE is based on the perturbation with respect to $J_{\rm sd}$, let us first consider the unperturbed system ($J_{\rm sd}=0$) where there is no interaction between $F$ and $P$. For the moment, we focus on the ferromagnet $F$. Under a uniform magnetic field $\bmH_0= H_0 \widehat{\bm z}$, the equilibrium spin $\bmS_{\rm eq}= S_{\rm eq} \widehat{\bm z}$ is determined by the condition $\bmH_{\rm eff}= {\bm 0}$, which yields the mean-field equation for $S_{\rm eq}$: 
%%%
\begin{equation}
  H_0 = \mathfrak{h}_0 (a_{\rm GL} S_{\rm eq}+ b_{\rm GL} S_{\rm eq}^3).
  \label{eq:Seq01}
\end{equation}
%%%
Therefore, in the limit of negligibly small external field $H_0 \approx 0$, the equilibrium spin in $F$, or the magnetization, is given by
%%%
\begin{equation}
  S_{\rm eq} = \sqrt{\frac{|a_{\rm GL}|}{b_{\rm GL}}} \propto \sqrt{T_{\rm c}- T}. 
\end{equation}
%%%
Now we consider the low-energy dynamics of $\bmS$, or the spin-wave excitation, by introducing the decomposition, 
%%%
\begin{equation}
  \bmS = \bmS_{\rm eq} + \delta \bmS,
  \label{eq:dS01}
\end{equation}
%%%
where $\delta \bmS$ represents a fluctuation about $\bmS_{\rm eq}$. Let $\bmS_{\bmq,\omega}$ be the Fourier transform of $\bmS(\bmr,t)$, 
%%%
\begin{equation}
  \bmS(\bmr,t) = \frac{1}{\sqrt{V}} \sum_\bmq \int_\omega \bmS_{\bmq}(\omega) e^{\ui \bmq \cdot \bmr- \ui \omega t}, 
\end{equation}
%%%
where $V$ is the system volume, and we have introduced the shorthand notation $\int_\omega= \int_{-\infty}^\infty \frac{d \omega}{2 \pi}$. Introducing $S^{\pm}= S^x \pm \ui S^y$ and $\xi^{\pm}= \xi^x \pm \ui \xi^y$, the transverse component of the TDGL equation~(\ref{eq:TDGL01}), which is linearized with respect to $\delta \bmS$, becomes
%%%
\begin{equation}
  (\omega- \omega_\bmq + \ui \Gamma^{+-}_{\bmq,{\rm eff}}) \delta S^-_{\bmq}(\omega)
  = \ui \xi^-_{\bmq}(\omega),
\end{equation}
%%%
where
%%%
\begin{equation}
  \omega_\bmq= \gamma (H_0 + \mathfrak{h}_0 S_{\rm eq}c_{\rm GL} q^2) 
\end{equation}
%%%
is the spin-wave resonance frequency. Besides, 
%%%
\begin{equation}
  %\Gamma^{+-}_{\bmq,{\rm eff}}= 
  \Gamma^{+-}_{\bmq,{\rm eff}}=
  \Gamma_{\bmq} (a_{\rm GL}+ b_{\rm GL} S_{\rm eq}^2+ c_{\rm GL}q^2)
  \approx \Gamma_{\bmq} (a_{\rm GL}+ b_{\rm GL} S_{\rm eq}^2) 
\end{equation}
is the effective damping constant of the transverse fluctuation, where we have introduced the notation 
%%%
\begin{equation}
  \Gamma_\bmq = \Gamma_0+ D_0q^2. 
\end{equation}
%%%

 Turning to the paramagnetic metal $P$, we introduce a decomposition similar to equation~(\ref{eq:dS01}):
%%%
\begin{equation}
  \bmsig = \bmsig_{\rm eq} + \delta \bmsig,
  \label{eq:dsig01}
\end{equation}
%%%
where the equilibrium value of the itinerant spin is given by 
%%%
\begin{equation}
  \bmsig_{\rm eq} = J_{\rm sd} \chi_P S_{\rm eq} \widehat{\bm z}. 
  \label{eq:dsigEQ01}
\end{equation}
%%%
Then, going into the Fourier space, the transverse component of the Bloch equation~(\ref{eq:Bloch01}) is represented as 
%%%
\begin{equation}
  (\omega + \ui \tau_P^{-1}) \delta \sigma^-_{\bmk}(\omega)
  = \ui \zeta^-_{\bmk}(\omega), 
\end{equation} 
%%%
where $\sigma^\pm = \sigma^x \pm \ui \sigma^y$ and $\zeta^\pm = \zeta^x \pm \ui \zeta^y$ as before.

%%%%%%%%%%%%%%%%%%%%%%%%%%%%%%%%%%%%%%%%%%%%%%%%%%%%%%%
\section{Local spin injection process} \label{Sec:III}
%%%%%%%%%%%%%%%%%%%%%%%%%%%%%%%%%%%%%%%%%%%%%%%%%%%%%%%
To discuss the longitudinal SSE in the present system, we first consider the effects of $s$-$d$ interaction at the $F$/$P$ interface under the temperature bias $\Delta T= T_F - T_P$. Since the interface breaks the translational symmetry of the system, the $s$-$d$ interaction mixes the wavenumbers of $\delta \bmS_\bmq$ and $\delta \bmsig_\bmk$. In the presence of the interfacial $s$-$d$ interaction, using the Fourier representation and performing a straightforward but a slightly tedious calculation, we obtain the following equations: 
%%%
\begin{eqnarray}
  (\omega- \omega_\bmq + \ui \Gamma^{+-}_{\bmq,{\rm eff}}) \delta S^-_{\bmq}(\omega)
  + \frac{J_{\rm sd}}{\hbar} S_{\rm eq}
  \sum_{\bmk'} \frac{\widetilde{\rho}_{\bmq-\bmk'}}{V} \sigma^-_{\bmk'}(\omega)
  &=& \ui \xi^-_{\bmq}(\omega), \label{eq:DSminus01}\\
  (\omega + \ui \tau_P^{-1}) \delta \sigma^-_{\bmk}(\omega)
  - \ui \frac{\chi_P}{\tau_P} J_{\rm sd}
  \sum_{\bmq'}\frac{\widetilde{\rho}_{\bmk-\bmq'}}{V} \delta S^-_{\bmq'}(\omega)
  &=& \ui \zeta^-_{\bmk}(\omega),  \label{eq:Dsigminus01} 
\end{eqnarray}
%%%
where we have introduced the Fourier representation of $\widetilde{\rho}$ as  
%%%
\begin{equation}
  \widetilde{\rho}(\bmr) = \frac{1}{V} \sum_\bmK \widetilde{\rho}_\bmK e^{\ui \bmK \cdot \bmr}. 
\end{equation}
%%%
Similarly, the equations for the $+$ branch is given by 
%%%
\begin{eqnarray}
  (\omega+ \omega_\bmq + \ui \Gamma^{+-}_{\bmq,{\rm eff}}) \delta S^+_{\bmq}(\omega)
  - \frac{J_{\rm sd}}{\hbar} S_{\rm eq}
  \sum_{\bmk'} \frac{\widetilde{\rho}_{\bmq-\bmk'}}{V} \sigma^+_{\bmk'}
  &=& \ui \xi^+_{\bmq}(\omega), \label{eq:DSplus01}\\
  (\omega + \ui \tau_P^{-1}) \delta \sigma^+_{\bmk}(\omega)
  - \ui \frac{\chi_P}{\tau_P} J_{\rm sd}
  \sum_{\bmq'} \frac{\widetilde{\rho}_{\bmk-\bmq'}}{V} \delta S^+_{\bmq'}(\omega)
  &=& \ui \zeta^+_{\bmk}(\omega),   \label{eq:Dsigplus01}
\end{eqnarray}
%%%

We are in a position to calculate the SSE in this system. We define the spin current injected from $F$ to $P$ by the SSE as follows: 
%%%
\begin{equation}
  j_{\rm SSE}= \frac{1}{A_{\rm contact}}
  \int \frac{d^3 r}{v_0} \frac{\partial}{\partial t} \bra \sigma^z(\bmr) \ket, 
\end{equation}
%%%
where $A_{\rm contact}$ is the contact area of the $F/P$ interface. The time derivative of $\sigma^z$ can be calculated from the $z$-component of the Bloch equation~(\ref{eq:Bloch01}), and using Fourier representation we obtain 
%%%
\begin{equation}
  j_{\rm SSE} = \frac{J_{\rm sd}}{A_{\rm contact}\hbar v_0} \sum_{\bmq,\bmk}
  \frac{\widetilde{\rho}_{\bmk- \bmq}}{V} 
  \int_\omega {\rm Im} \bra \bra \delta S^-_{\bmq}(\omega) \delta \sigma^+_{-\bmk}(-\omega) \ket \ket,
  \label{eq:ISSE01}
  \end{equation}
%%%
where we have introduced the notation $\bra \delta S^-_{\bmq}(\omega) \delta \sigma^+_{-\bmk}(\omega') \ket = 2 \pi \delta (\omega+ \omega') \bra \bra \delta S^-_{\bmq}(\omega) \delta \sigma^+_{-\bmk}(-\omega) \ket \ket$ for the correlation in the frequency space. Now our remaining task is to evaluate the transverse correlation $\bra \bra \delta S^-_{\bmq}(\omega) \delta \sigma^+_{-\bmk}(-\omega) \ket \ket$. Using perturbation approach to the coupled equations (\ref{eq:DSminus01}) and (\ref{eq:Dsigminus01}) with respect to $J_{\rm sd}$, $\delta S^-_{\bmq}(\omega)$ is solved to be 
%%%
\begin{equation}
  \delta S^-_{\bmq}(\omega) =
  G^-_{\bmq}(\omega)  \ui \xi^-_{\bmq}(\omega) 
  - \frac{J_{\rm sd}}{\hbar} S_{\rm eq} G^-_{\bmq}(\omega)
  \sum_{\bmk'} \frac{\widetilde{\rho}_{\bmq-\bmk'}}{V} 
  g_{\bmk'}(\omega) \ui \zeta^-_{\bmk'}(\omega) ,
  \label{eq:DSminus02}
\end{equation}
%%%
where we have defined $G^-_{\bmq}(\omega)= (\omega- \omega_\bmq + \ui \Gamma^{+-}_{\bmq,{\rm eff}})^{-1}$ and $g_{\bmk}(\omega)= (\omega + \ui \tau_P^{-1})^{-1}$. In a similar manner, from the coupled equations (\ref{eq:DSplus01}) and (\ref{eq:Dsigplus01}), we obtain 
%%%
\begin{eqnarray}
  \delta \sigma^+_{-\bmk}(-\omega) &=&
  g_{-\bmk}(-\omega)  \zeta^+_{-\bmk}(-\omega) \nonumber \\
  &+& \ui \frac{J_{\rm sd} \chi_P}{\tau_P} g_{-\bmk}(-\omega)
  \sum_{\bmq'} \frac{\widetilde{\rho}_{-\bmk+\bmq'}}{V} 
  G^+_{-\bmq'}(-\omega) \ui \xi^+_{-\bmq'}(-\omega) , 
  \label{eq:Dsigplus02}
\end{eqnarray}
%%%
where $G^+_{\bmq}(\omega)= (\omega+ \omega_\bmq + \ui \Gamma^{+-}_{\bmq,{\rm eff}})^{-1}$. 

We substitute equations (\ref{eq:DSminus02}) and (\ref{eq:Dsigplus02}) into equation~(\ref{eq:ISSE01}) to calculate $j_{\rm SSE}$. Recalling that there is no cross correlation between the two noises ${\bm \xi}$ and ${\bm \zeta}$, and using that both $\bra \bra \xi^-_\bmq(\omega) \xi^+_{\bmq'}(-\omega) \ket \ket$ and $\bra \bra \zeta^-_\bmq(\omega) \zeta^+_{\bmq'}(-\omega) \ket \ket$ are proportional to $\delta_{\bmq,-\bmq'}$, we find that the injected spin current can be divided into two contributions:
%%%
\begin{equation}
  j_{\rm SSE} = j^{\rm pump}_{\rm SSE}- j^{\rm back}_{\rm SSE},
\end{equation}
%%%
where the two terms $j^{\rm pump}_{\rm SSE}$ and $j^{\rm back}_{\rm SSE}$ are defined by 
%%%
\begin{eqnarray}
  j_{\rm SSE}^{\rm pump} &=&
  -\sum_{\bmq,\bmk} \frac{J_{\rm sd}^2|\widetilde{\rho}_{\bmk-\bmq}|^2}{A_{\rm contact}\hbar v_0 V^2} 
  \int_\omega |G^-_\bmq(\omega)|^2 |g_\bmk(\omega)|^2 \frac{\omega \chi_P }{\tau_P}
  \bra \bra \xi^-_\bmq(\omega) \xi^+_{-\bmq}(-\omega) \ket \ket , \nonumber \\
  \\ 
  j_{\rm SSE}^{\rm back} &=&
  - \sum_{\bmq,\bmk} \frac{J_{\rm sd}^2|\widetilde{\rho}_{\bmk-\bmq}|^2}{A_{\rm contact}\hbar v_0 V^2} 
  \int_\omega |G^-_\bmq(\omega)|^2 |g_\bmk(\omega)|^2
  \frac{S_{\rm eq} \Gamma^{+-}_{\bmq,{\rm eff}}}{\hbar}
  \bra \bra \zeta^-_\bmk(\omega) \zeta^+_{-\bmk}(-\omega) \ket \ket, \nonumber \\
\end{eqnarray}
%%%
and we have used the properties $G_{-\bmq}^+(-\omega)= -[G_{\bmq}^-(\omega)]^*$ and $g_{-\bmk}(-\omega)= -[g_{\bmk}(\omega)]^*$. We then recall the fluctuation-dissipation relations (\ref{eq:FDxi01}) and (\ref{eq:FDzeta01}), which in the momentum space become $\bra \bra \xi^-_\bmq(\omega) \xi^+_{-\bmq}(-\omega) \ket \ket= 4 \kB T_F \Gamma_\bmq/\eps_0$ and $\bra \bra \zeta^-_\bmk(\omega) \zeta^+_{-\bmk}(-\omega) \ket \ket= 4 \kB T_P \chi_P v_0/\tau_P$. The integral over the frequency $\omega$ can be done by picking up the magnon pole at $\omega= \omega_\bmq+ \ui \Gamma^{+-}_{\bmq,{\rm eff}}$, yielding  
%%%
\begin{equation}
  j_{\rm SSE} = -\frac{2 N_{\rm int}J^2_{\rm sd} \chi_P \tau_P}{A_{\rm contact}\hbar v_0 N_F N_P}
  \sum_{\bmq,\bmk}  
  \left\{
  \frac{\Gamma_\bmq \omega_\bmq}{\Gamma^{+-}_{\bmq,{\rm eff}} \eps_0} 
  \kB T_F
  - \frac{S_{\rm eq}}{\hbar} v_0 \kB T_P
  \right\}, 
\end{equation}
%%%
where $N_F$ and $N_P$ are respectively the number of lattice sites in $F$ and $P$, and we used an approximation $\omega_\bmq \tau_P \ll 1$. Also, we assumed a diffuse-scattering interface and hence used $|\widetilde{\rho}_{\bmk- \bmq}|^2 \simeq N_{\rm int}v_0^2$ with $N_{\rm int}$ being the number of localized spins at the $F/P$ interface. A further simplification can be made by the relation: 
%%%
\begin{equation}
  \Gamma_\bmq \omega_\bmq =
  \frac{\eps_0 v_0}{\hbar} \Gamma^{+-}_{\bmq,{\rm eff}} S_{\rm eq}
\end{equation}
%%%
which can be proven using the mean-field equation (\ref{eq:Seq01}) for $S_{\rm eq}$. Using that the momentum sum approximately returns unity, $N_F^{-1}N_P^{-1} \sum_{\bmq,\bmk} \simeq 1$, we finally obtain
%%%
\begin{equation}
  j_{\rm SSE} = -\frac{2 N_{\rm int} J^2_{\rm sd} \chi_P \tau_P}{A_{\rm contact}\hbar^2 }
  S_{\rm eq} \kB \Delta T, 
  \label{eq:ISSE02}
\end{equation}
%%%
where $\Delta T = T_F - T_P$ as stated at the beginning of this section.  

%%%%%%%%%%%%%%%%%%%%%%%%%%%%%%%%%%%%%%%%%%%%%%%%%%%%%%%%%%%%%%%%%%%%%%%%%
\section{Effects of spin diffusion inside the ferromagnet} \label{Sec:IV}
%%%%%%%%%%%%%%%%%%%%%%%%%%%%%%%%%%%%%%%%%%%%%%%%%%%%%%%%%%%%%%%%%%%%%%%%%
So far, we have discussed the SSE in a simple ferromagnet near $T_{\rm c}$, by focusing on the local spin injection process. In other words, the quantity we have just calculated corresponds to the interfacial spin conductance $G_s$ across $F/P$. This can be seen by rewriting equation~(\ref{eq:ISSE02}) as follows:
%%%
\begin{eqnarray}
  j_{\rm SSE} &=& -G_{s} \delta n_{\rm sw} , \label{eq:Gs01}\\
  G_{s} &=& \frac{2 N_{\rm int} J^2_{\rm sd} \chi_P \tau_P \kB T v_0}
  {A_{\rm contact} \hbar^2 } S_{\rm eq} 
\end{eqnarray}
%%%
where $\delta n_{\rm sw}= v_0^{-1}\Delta T/T$ has the meaning of deviation of the spin-wave density from its equilibrium value, and the negative sign before $G_s$ arises from defining the positive direction of $j_{\rm SSE}$. The above expression means that if there is a nonequilibrium spin-wave density $\delta n_{\rm sw}$, there arises a finite spin injection.

Now it is our common wisdom through the examination of the ferromagnet-thickness dependence of the longitudinal SSE~\cite{Kehlberger15} and its theoretical interpretation~\cite{Rezende14}, that a proper description of the longitudinal SSE requires the information on the spin diffusion inside the ferromagnet, which is represented by the following transport equation~\cite{ZhangZhang12}: 
%%%
\begin{equation}
  J_{\rm sw} = -D_{\rm sw} \nabla \delta n_{\rm sw} - {\cal S}_{\rm sw} \nabla T,
  \label{eq:transport01}
\end{equation}
%%%
where $J_{\rm sw}$ is the spin-wave spin current, and the two coefficients $D_{\rm sw}$ and ${\cal S}_{\rm sw}$ are defined phenomenologically by the above equation. Such a transport equation is recently discussed in analyzing the nonlocal spin transport in a lateral YIG/Pt system~\cite{Shan16}. Since the spin diffusion is known to show an anomaly concomitant with the critical slowing down near $T_{\rm c}$, it is of importance to investigate the effects of spin diffusion on the longitudinal SSE. 

Following Refs.~\cite{ZhangZhang12,Rezende14}, after considering the effects of magnon diffusion through the ferromagnet, the spin current $J_{\rm SSE}$ injected by the longitudinal SSE now takes the form, 
%%%
\begin{equation}
  J_{\rm SSE} = G_{s} 
  \frac{\cosh(L/\Lambda_{\rm sw})-1}{\sinh(L/\Lambda_{\rm sw})}
  \left (-\frac{\Lambda_{\rm sw} {\cal S}_{\rm sw}}{D_{\rm sw}} \nabla T \right), 
\end{equation}
%%%
where $L$ is the thickness of $F$ (see Figure~\ref{fig1_adachi}). The crucial finding is that the two transport coefficients $D_{\rm sw}$ and ${\cal S}_{\rm sw}$ appear in a pair as ${\cal S}_{\rm sw}/D_{\rm sw}$, for which the singularities in $D_{\rm sw}$ and ${\cal S}_{\rm sw}$ cancel, leaving only a regular behavior. Below, we show that there appears no singularity in ${\cal S}_{\rm sw}/D_{\rm sw}$ as well as in $\Lambda_{\rm sw}$.

Let us first discuss how the spin diffusion length behaves near $T_{\rm c}$ in the present model. For this purpose, we consider the $z$-component of the TDGL equation (\ref{eq:TDGL01}) with no noise term in the absence of $J_{\rm sd}$:
%%%
\begin{equation}
  \frac{\partial}{\partial t} \delta S^z 
  =
  \Big( - D^z_{\rm eff} \nabla^2 - \Gamma^z_{\rm eff} \Big) \delta S^z, 
  \label{eq:dSz01}
\end{equation}
%%%
where the two coefficients $D^z_{\rm eff}$ and $\Gamma^z_{\rm eff}$ have the same renormalization factor as
%%%
\begin{eqnarray}
  D^z_{\rm eff} &=& D_0 (a_{\rm GL}+ 3 b_{\rm GL}S^2_{\rm eq}), \label{eq:Dz01}\\
  \Gamma^z_{\rm eff} &=& \Gamma_0 (a_{\rm GL}+ 3 b_{\rm GL}S^2_{\rm eq}) \label{eq:Gammaz01}. 
\end{eqnarray}
Equation~(\ref{eq:dSz01}) has the form of spin diffusion equation, where $D^z_{\rm eff}$ has the meaning of the spin diffusion coefficient and $\Gamma^z_{\rm eff}$ has the meaning of the inverse spin relaxation time. Temperature dependence of these two coefficients are already studied in Ref.~\cite{Ohnuma14} (see Fig.~6(a) therein), and we see that these two coefficients $D^z_{\rm eff}$ and $\Gamma^z_{\rm eff}$ show a critical slowing down, i.e., $D^z_{\rm eff}, \Gamma^z_{\rm eff} \propto T_{\rm c}-T$, upon approaching $T \to T_{\rm c}$ from the ordered state. Note that this result is consistent with the conventional theory~\cite{Mori62}. From equations (\ref{eq:Dz01}) and (\ref{eq:Gammaz01}), the spin diffusion length is identified as
%%%
\begin{equation}
  \Lambda_{\rm sw} = \sqrt{\frac{D^z_{\rm eff}}{\Gamma^z_{\rm eff}}} =  \sqrt{\frac{D_0}{\Gamma_0}}, 
\end{equation}
%%%
which means that the spin diffusion length is not affected by the critical slowing down of each coefficient $D^z_{\rm eff}$ or $\Gamma^z_{\rm eff}$. 

Next, following Luttinger's derivation of Einstein relation between the diffusion coefficient and the conductivity~\cite{Luttinger64}, we argue that the coefficients ${\cal S}_{\rm sw}$ is proportional to the spin diffusion coefficient $D_{\rm sw}$, leaving the combination ${\cal S}_{\rm sw}/D_{\rm sw}$ nonsingular. From the transport equation~(\ref{eq:transport01}) we obtain in the momentum space, 
%%%
\begin{equation}
  {\bm j}_{{\rm sw}, \bmq}= -\ui \bmq D_{\rm sw} \delta n_{{\rm sw},\bmq} - \ui \bmq {\cal S}_{\rm sw} T_\bmq .
\end{equation}
%%%
We also consider the spin continuity equation, which is approximately given by
%%%
\begin{equation}
  s \delta n_{{\rm sw}, \bmq} + \ui \bmq \cdot {\bm j}_{{\rm sw},\bmq} = 0, 
\end{equation}
%%%
where a time dependence of the type $n_{{\rm sw},\bmq} \propto e^{st}$ with a small positive constant $s$ is assumed as in Ref.~\cite{Luttinger64}. From these two equations, we obtain $\delta n_{{\rm sw}, \bmq}= {q^2 {\cal S}_{\rm sw} T_{\bmq}}/({s + D_{\rm sw}q^2})$, which in the relevant ``slow limit'' $D_{\rm sw} q^2 \gg s$ becomes 
%%%
\begin{equation}
  \delta n_{{\rm sw}, \bmq}= \frac{{\cal S}_{\rm sw} }{D_{\rm sw}}T_{\bmq}. 
\end{equation}
%%%
The both sides of the above equation in the ``slow limit'' are equilibrium quantities, such that the ratio ${\cal S}_{\rm sw}/D_{\rm sw}$ can be expressed in terms of equilibrium property of the system. Because the singularity concomitant with the critical slowing down has intrinsically a dynamic nature, we argue that no singularity appears in the ratio ${\cal S}_{\rm sw}/D_{\rm sw}$. Therefore, the result obtained in the previous section that the SSE signal $J_{\rm SSE}$ scales with the magnetization near $T_{\rm c}$, i.e., $J_{\rm SSE} \propto (T_{\rm c}-T)^{1/2}$, remains valid even if the effects of spin diffusion through the ferromagnet is taken into account. 

%%%%%%%%%%%%%%%%%%%%%%%%%%%%%%%%%%%%%%%%%%%%%%%%%
\section{Discussion and conclusion} \label{Sec:V}
%%%%%%%%%%%%%%%%%%%%%%%%%%%%%%%%%%%%%%%%%%%%%%%%%
In this work, the longitudinal SSE in a simple ferromagnet near $T_{\rm c}$ has been examined on the basis of the TDGL model. It was found analytically that the SSE shows a power law behavior $J_{\rm SSE} \sim (T_{\rm c}-T)^{1/2}$, and that the conclusion remains unchanged even if we take account of the effects of spin diffusion inside the ferromagnet. Interestingly, the present analytical result obtained from the TDGL model is consistent with the previous atomistic numerical simulation of Heisenberg Hamiltonian~\cite{Barker16}. While our conclusion differs from the experiment studying the longitudinal SSE in a YIG/Pt system near $T_{\rm c}$~\cite{Uchida14}, we think that our result suggests the importance of considering the ferrimagnetic nature of YIG in order to account for the experiment. Besides, it may be important to take care of the intrinsic magnetic surface anisotropy in a YIG/Pt system~\cite{Uchida15}, since the surface anisotropy may substantially reduce the SSE signal from that without the anisotropy when the magnetization is small. 

Before conclusion, we would like to propose an experiment which can test our theoretical result. Europium oxide (EuO) is an idealistic ferromagnetic semiconductor, with a bandgap around $1.2$~eV and with the Curie temperature $T_{\rm c} = 69.3 $~K~\cite{Coey-text}. At low temperatures, this magnet and related magnet EuS ($T_{\rm c} =16.5$~K) may be modeled as a simple insulating ferromagnet. Therefore, it is tempting to study the SSE in EuO/Pt and EuS/Pt systems in order to see if the SSE near $T_{\rm c}$ shows a power law behavior $\sim (T_{\rm c}-T)^{1/2}$ as predicted in this work. 

To conclude, we have examined the longitudinal SSE in a simple ferromagnet near $T_{\rm c}$ by a TDGL approach. We found that the SSE signal is proportional to the magnetization near $T_{\rm c}$ as $J_{\rm SSE} \sim (T_{\rm c}-T)^{1/2}$. Our analytical result on the basis of TDGL model is consistent with the previous numerical simulation which employs quite different model of atomistic Heisenberg Hamiltonian~\cite{Barker16}. Since the longitudinal SSE near $T_{\rm c}$ has only been studied in a rather complicated magnetic system YIG/Pt~\cite{Uchida14}, we hope that the present result is tested experimentally in a more simple system such as EuO/Pt or EuS/Pt. 

\section*{Acknowledgments}
This work was financially supported by JSPS KAKENHI Grant No. 15K05151. 

\section*{References}
%%%%%%%%%%%%%%%%%%%%%%%%%%%%%%%%%%

\end{document}